\documentclass[
preprint,
amsmath,amssymb,
aps,
prl,
superscriptaddress
]{revtex4-1}

\usepackage{graphicx}
\usepackage{dcolumn}
\usepackage{bm} 
\usepackage{stmaryrd}
\usepackage{pxfonts}
\usepackage{txfonts}
\usepackage{float}
\usepackage{ulem}
\usepackage{amssymb}
\usepackage{amsmath}
\usepackage[dvipsnames]{xcolor}
\usepackage{textcomp}
\usepackage{color}
\usepackage{multirow}
\usepackage{afterpage}

\begin{document}

\title{Low-frequency quantum oscillations in LaRhIn$_5$: Dirac point  or nodal line?\footnote{ARISING FROM C. Guo et al. Nature Communications https://doi.org/10.1038/s41467-021-26450-1 (2021)
}}

\author{G.\,P.\,Mikitik}
\email[Corresponding author: ]{mikitik@ilt.kharkov.ua}
\affiliation{B. Verkin Institute for Low Temperature Physics and Engineering of National Academy of Sciences of Ukraine, Kharkiv 61103, Ukraine}
\author{Yu.\,V.\,Sharlai}
\affiliation{B. Verkin Institute for Low Temperature Physics and Engineering of National Academy of Sciences of Ukraine, Kharkiv 61103, Ukraine}
\affiliation{Institute of Low Temperature and Structure Research, Polish Academy of  Sciences, 50-422 Wroc{\l}aw, Poland}

\maketitle
\thispagestyle{empty}

In the recent paper \cite{guo}, a new method based on measuring a temperature correction to a quantum-oscillation frequency was proposed to study an energy-band dispersion of charge
carriers in small Fermi surface (FS) pockets of crystals.
To illustrate their approach, Guo et al.\ \cite{guo} applied it to a number of materials and, in particular, to the multiband metal LaRhIn$_5$ which, apart from high-frequency oscillations associated with a large FS, also exhibits the oscillations with the low frequency $F\approx 7$ T. Although the method of Ref.~\cite{guo} really detects charge carriers with a linear dispersion, it  does not distinguish between the carriers near a Dirac point and near a nodal line, since all such quasiparticles disperse linearly. Here we ask what is the nature of the carriers associated with the frequency $F$ in LaRhIn$_5$ and call attention to the puzzling origin of this frequency.

\begin{figure}[tbp] 
 \centering  \vspace{+9 pt}
\includegraphics[scale=1]{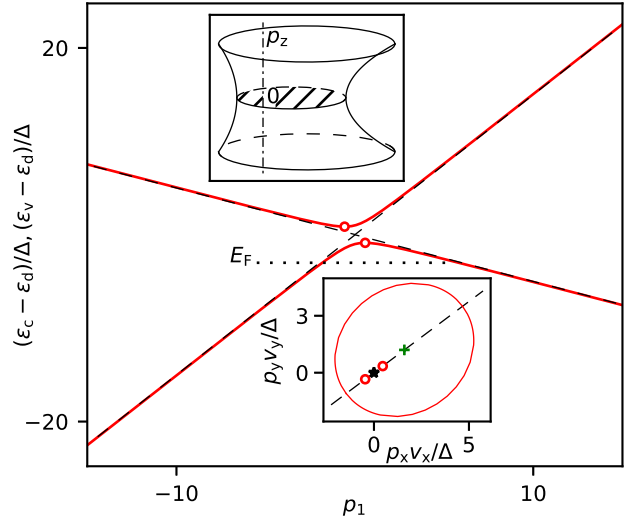}
\caption{\label{fig1} \textbf{The energy bands $\varepsilon_{\rm c}({\bf p})$ and $\varepsilon_{\rm v}({\bf p})$, Eq.~(\ref{1}), in the vicinity of their nodal line in the plane $p_z=0$ perpendicular to the line.} The red solid and black dashed lines show the bands with and without the spin-orbit interaction, respectively. The red circles mark the minimum of $\varepsilon_{\rm c}({\bf p})$ and the maximum of $\varepsilon_{\rm v}({\bf p})$ in the plane. The minimal indirect gap $2\Delta_{\rm min}= 2\Delta(1-\tilde a_{\perp}^2)^{1/2}$ determined by these two points is less than $2\Delta$, the spin-orbit gap at ${\bf p}=0$. Here $p_{1}\equiv (a_xp_x+a_yp_y)/(\tilde a_{\perp}\Delta)$ is the dimensionless quasimomentum measured along the vector $(\tilde a_x,\tilde a_y)$ in the plane with the coordinates $p_xv_x/\Delta$ and $p_yv_y/\Delta$;  $\tilde a_i\equiv a_i/v_i$, and $\tilde a_{\perp}\equiv(\tilde a_x^2+\tilde a_y^2)^{1/2}$. The dotted line indicates the Fermi level $E_{\rm F}$. Upper inset: The Fermi surface enclosing the nodal line (the dash-dotted line) at $(E_{\rm F}-\varepsilon_{\rm d})b<0$. Lower inset: The cross section (ellipse) of the Fermi surface on the plane $p_z=0$. The black dashed line marks the direction along which the bands are shown in the main panel.
The black asterisk and green cross mark the point ${\bf p}=0$ and the center of the ellipse, respectively.
 } \end{figure}

Many years ago \cite{m-sh04}, we argued that the $7$ T frequency is due to the minimal cross section of a FS surrounding a nodal line in LaRhIn$_5$, whereas Guo et al.~\cite{guo} now relate this frequency with a cross section of a FS pocket enclosing a Dirac point. Below we show that the main experimental result of Ref.\ \cite{guo} does not contradict our assumption of the nodal line in LaRhIn$_5$.
The degeneracy of two bands $\varepsilon_{\rm c}({\bf p})$ and $\varepsilon_{\rm v}({\bf p})$ along a nodal line, strictly speaking, occurs in LaRhIn$_5$ only when neglecting a weak spin-orbit interaction. Consider now these bands in the vicinity of some point ${\bf p}_0$ of the line, taking into account this interaction  \cite{jetp,m-sh19} (Fig.~\ref{fig1}),
\begin{equation}\label{1}
\varepsilon_{\rm c,v}({\bf p})\!=\!\varepsilon_{\rm d}+{\bf a}{\bf p}
+bp_z^2\pm \!\!\sqrt{\Delta^2\!+\!(v_xp_x)^2\!+\!(v_yp_y)^2}
\end{equation}
where $\Delta\equiv \Delta({\bf p}_0)$ is half of the spin-orbit gap at the point ${\bf p}_0$, $\varepsilon_{\rm d}$ is the band-degeneracy energy at this point in absence of the spin-orbit coupling (i.e., when $\Delta=0$), $v_x$, $v_y$, ${\bf a}=(a_x,a_y,a_z)$, and $b$ are constant parameters, the quasimomentum ${\bf p}$ is measured from ${\bf p}_0$, the $p_z$ axis coincides with the tangent to the band-contact line at this point, and the $p_x$, $p_y$ axes are chosen in such a way that the quadratic form  under the square root is diagonal. Let the magnetic field be directed along $p_z$. The cross section of a FS surrounding the nodal line at the plane $p_z=$ constant is a closed curve (an ellipse) only if $\tilde a_{\perp}^2\equiv (a_x/v_x)^2+(a_y/v_y)^2 <1$. The parameter $\tilde a_{\perp}$ characterizes the tilt of the spectrum at constant $p_z$, and  $\tilde a_{\perp}\neq 0$ for all real situations. If the cross-sectional area at $p_z=0$ is extremal with respect to $p_z$, then $a_z=0$, and the term $bp_z^2$ is taken into account in Eq.~(\ref{1}).

The temperature dependence of the quantum-oscillation frequency $F$ looks as follows \cite{guo}:
 \begin{eqnarray}\label{2}
F(E_{\rm F})=F_0-\theta\frac{(\pi k_{\rm B}T)^2}{F_0\beta^2}
 \end{eqnarray}
where $E_{\rm F}$ is the Fermi energy,  $F_0$ is the frequency of these oscillations at zero temperature, $\beta=e\hbar/2m_{\rm c}$,
$m_{\rm c}$ is the cyclotron mass, and $\theta=1/16$ for a band with a linear dispersion. With Eq.~(\ref{1}) and formulas of Guo et al.~\cite{guo} for $\theta$, we arrive at
 \begin{eqnarray}\label{3}
 \theta&=&\frac{1}{16}\left(1-\frac{\Delta_{\rm min}^2}{(E_{\rm F}-\varepsilon_{\rm d})^2}\right)
 \end{eqnarray}
where $\Delta_{\rm min}=\Delta(1-\tilde a_{\perp}^2)^{1/2}$ is the minimal indirect half-gap in the plane $p_z=0$ (Fig.~\ref{fig1}). In Ref.~\cite{guo}, the simplified spectrum with $a_x=a_y=\tilde a_{\perp}=0$ was implied, and it was concluded that the value $\theta=1/16$, which was experimentally obtained for LaRhIn$_5$,
can occur only if the direct spin-orbit gap is perturbatively small,  $\Delta^2/(E_{\rm F}-\varepsilon_{\rm d})^2 \ll 1$. On the other hand, the band-structure calculations \cite{guo} revealed that this ratio, in general, is not very small for LaRhIn$_5$, and Guo et al.\ ascribed the frequency $F$ to a cross section passing through a Dirac point (when $\Delta\equiv 0$), excluding the case of the nodal line from their  consideration. However, Eq.~(\ref{3}) demonstrates that
the perturbatively small $\Delta$ is not necessary to obtain  $\theta\approx 1/16$. It is sufficient if only the indirect spin-orbit  gap in the plane of the extremal cross section of the FS is small, $\Delta_{\rm min}^2/(E_{\rm F}-\varepsilon_{\rm d})^2 \ll 1$, and so a nodal line can lead to $\theta\approx 1/16$ even though the spin-orbit coupling is not perturbatively weak in LnRhIn$_5$.

\begin{figure}[tbp] 
 \centering  \vspace{+9 pt}
\includegraphics[scale=1]{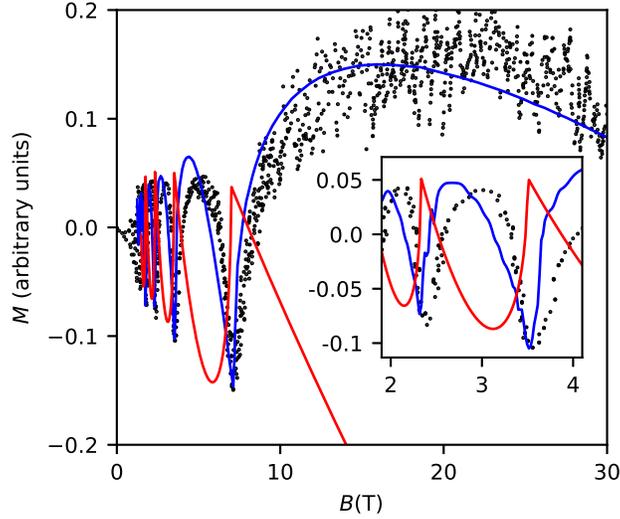}
\caption{\label{fig2} \textbf{Magnetization of LaRhIn$_5$}. The dots are the experimental data \cite{goodrich}, the blue line shows the magnetization \cite{m-sh04} produced by the nodal line in Fig.~\ref{fig1}, whereas the red line is the magnetization of the Dirac pocket, with the background term $\chi_0B=-0.7Cg(1)B$ being added to Eq.~(\ref{4}). This term is determined by the charge carriers that are far away from the point ${\bf p}=0$ \cite{m-sh04}.
 The inset is a zoom into the region $2\le B\le 4$ T.
 } \end{figure}   

It was shown earlier \cite{m-sh04} that the experimental dependence of the longitudinal magnetization of LaRhIn$_5$ on the magnetic induction $B$ \cite{goodrich} can be explained if a nodal line penetrates the minimal cross section of the FS in this material (Fig.~\ref{fig2}). Let us now discuss the case of a FS  pocket enclosing the Dirac point assumed by Guo et al.\ \cite{guo}.  A formula for the magnetization of such a pocket with a linear dispersion of its charge carriers was derived many years ago \cite{m-sh}, and a convenient representation \cite{m-sh16,m-sh19} of this formula reads:
\begin{eqnarray}\label{4}
M= C F g(u)
 \end{eqnarray}
where the positive coefficient $C$ depends on Dirac-spectrum parameters, $u\equiv F/B$, and $g(u)$ is a universal function independent of any parameters. The magnetization calculated with Eq.~(\ref{4}) at $F=7$ T is shown in Fig.~\ref{fig2}. The value of $C$ is chosen in such a way that the calculated amplitude of the
oscillations agrees with the experimental data. Figure 2 reveals a qualitative disagreement between this theoretical curve and the data. The theoretical curve (which is the same for electron and hole Dirac pockets) exhibits sharp peaks, whereas the experimental data reveal sharp troughs. Moreover, the behavior of the magnetization at $B>F$  essentially deviates from the experimental dependence $M(B)$. In other words, the assumption that the low-frequency oscillations are determined by a Dirac pocket is incompatible with the data \cite{goodrich} on the magnetization of LaRhIn$_5$. On the other hand, Supplementary Fig.~4 of Guo et al.~\cite{guo} shows that the frequency $F=7$ T of the Shubnikov--de Haas oscillations is practically independent of the direction of the magnetic field. This result is inconsistent with the nodal-line assumption, which leads to $F(\psi)\sim  1/\cos\psi$ where $\psi$ is the angle between $B$ and the line, and so Guo et al.\ \cite{guo} assumed the existence of the Dirac point. Thus, at present there is no self-consistent explanation of the $7$ T oscillations in LaRhIn$_5$.

The Fermi surface near the nodal line (upper inset in Fig.~\ref{fig1}) and the appropriate $M(B)$ (blue line in Fig.~\ref{fig2}) are shown for the case $(E_{\rm F}- \varepsilon_{\rm d})b<0$. If the small difference $E_{\rm F}-\varepsilon_{\rm d}$ changes its sign, the anisotropy of the frequency $F(\psi)$ noticeably decreases, and $M(B)$ resembles the red curve in Fig.~\ref{fig2} \cite{m-sh04}. Thus, the previously published data \cite{guo,goodrich} look as if they were obtained on crystals with slightly different $E_{\rm F}$ but with practically equal $|E_{\rm F}-\varepsilon_{\rm d}|$. This hypothesis can be verified, measuring both the Shubnikov--de Haas oscillations and the longitudinal magnetization $M(B)$ in one and the same sample. Then, in a sample with $M(B)$ like in Goodrich et al.~\cite{goodrich}, the dependence  $F(\psi)$ has to be strongly anisotropic, whereas in a sample with a weak dependence $F(\psi)$, $M(B)$ cannot exhibit the sharp troughs visible in the oscillations in Fig.~\ref{fig2}.

\textbf{Acknowledgements.}
Yu.V.S. acknowledges the Poland NCN program for scientists from Ukraine [R-2022/01/3/ST3/00083].


\begin{references}
	

\bibitem{guo} Guo, C. et al. Temperature dependence of quantum oscillations from non-parabolic dispersion. \textit{Nat. Commun.} \textbf{12}, 6213 (2021).


\bibitem{m-sh04} Mikitik, G. P. \& Sharlai, Yu. V. Berry phase and de Haas - van Alphen effect in LaRhIn$_5$. \textit{Phys. Rev. Lett.} \textbf{93}, 106403 (2004).


\bibitem{jetp} Mikitik, G.P. \& Sharlai Yu.V. Semiclassical energy levels of electrons in metals with band degeneracy lines.
    \textit{JETP} {\bf 87}, 747-755 (1998).


\bibitem{m-sh19} Mikitik, G. P. \& Sharlai, Yu. V. Magnetic Susceptibility of Topological Semimetals. \textit{J. Low Temp. Phys.} \textbf{197}, 272-309 (2019).

\bibitem{goodrich}  Goodrich, R.G. et al. Magnetization in the ultraquantum limit. \textit{Phys. Rev. Lett.} \textbf{89}, 026401  (2002).

\bibitem{m-sh} Mikitik, G.P., Sharlai, Yu.V. Field dependences of magnetic susceptibility of crystals under conditions of degeneracy of their electron energy bands.
       \textit{Low Temp. Phys.} {\bf 22}, 585-592 (1996).


\bibitem{m-sh16} Mikitik, G. P. \& Sharlai, Yu. V. Magnetic susceptibility of topological nodal
semimetals. \textit{Phys. Rev. B} \textbf{94}, 195123 (2016).


\end{references}
\end{document}